\documentclass[aps,prl,twocolumn,groupedaddress]{revtex4-2}
\usepackage{amsmath,amsthm,amssymb}
\usepackage{graphicx}

\begin{document}

\newcommand{\bi}[1]{\ensuremath{\boldsymbol{#1}}} 

\title{Quantized Hyperfine Field at an Implanted $\mu^+$ Site in
PrPb$_3$:\\ Interplay between Localized $f$ Electrons and
an Interstitial Charged Particle}

\author{T.~U.~Ito$^{1,2}$}\email[Contact author: ito.takashi15@jaea.go.jp]\ 
\author{W.~Higemoto$^2$} 
\author{K.~Ohishi$^{2}$}
\author{N.~Nishida$^1$} 
\author{R.~H.~Heffner$^{2,3}$}
\author{Y.~Aoki$^4$}
\author{A.~Amato$^5$} 
\author{T.~Onimaru$^6$}
\author{H.~S.~Suzuki$^7$}
\affiliation{$^1$Department of Physics, Tokyo Institute of Technology, Meguro,
 Tokyo 152-8551, Japan}
\affiliation{$^2$Advanced Science Research Center, Japan Atomic Energy
 Agency, Tokai, Ibaraki 319-1195, Japan}
\affiliation{$^3$Los Alamos National Laboratory, Los Alamos, New Mexico 87545, USA}
\affiliation{$^4$Department of Physics, Tokyo Metropolitan University, Hachioji, 
Tokyo 192-0397, Japan}
\affiliation{$^5$Laboratory for Muon-Spin Spectroscopy, Paul Scherrer
 Institute, CH-5232 Villigen PSI, Switzerland}
\affiliation{$^6$Department of Quantum Matter, ADSM, Hiroshima University, Higashi-Hiroshima, Hiroshima 739-8530, Japan}
\affiliation{$^7$National Institute for Materials Science, Tsukuba, Ibaraki
 305-0047, Japan}


\begin{abstract}
The local effect of an interstitial hydrogen-like particle on localized
 $f$ electrons was studied in PrPb$_3$ by means of $\mu^+$ spin
 rotation and relaxation. Spontaneous $\mu^+$ spin precession
 with harmonic frequencies was observed for the first time in
 $f$ electron compounds. We demonstrate that the signal is derived from
 a coupling between the $\mu^+$ spin and the hyperfine-enhanced
 nuclear spin of nearest neighbor (nn) $^{141}$Pr with Ising-like
 anisotropy. The signal also suggests a marked suppression of
 spin dynamics of the nn $^{141}$Pr in comparison with that of the
 bulk $^{141}$Pr. These facts strongly indicate modification of the 
 $f$ electronic state due to the interstitial charged particle.
\end{abstract}

\pacs{71.70.Ch, 76.75.+i}

\maketitle
The behavior of $^1$H impurities in rare-earth (RE) based compounds and
the influence on their surroundings are crucial issues for practical
materials science. 
$^1$H is the most common impurity which affects the quality of
samples. In fact, $^1$H is sometimes deliberately loaded into
a material to improve functions of the base
compound~\cite{huiberts96}, or to
store $^1$H itself at interstitial sites of the crystal~\cite{kadir97}. 
In RE-containing materials the electronic state of the RE ion should be
sensitive to the $^1$H
impurity because the positive charge of the $^1$H creates an additional
crystalline-electric-field (CEF) potential which affects electromagnetic
properties of the nearby RE ions.
A detailed understanding of the effects of $^1$H impurities in RE-based
compounds is therefore important for industrial applications
as well as for fundamental science. 

One of the most efficient techniques to investigate microscopic
properties of interstitial $^1$H is positive muon spin
rotation and relaxation ($\mu^+$SR), which probes the local magnetic environment
around an interstitial $\mu^+$, a light isotope of $^1$H.
In this letter, we report $\mu^+$SR measurements in PrPb$_3$ to clarify
the local effect of the interstitial $\mu^+$ on the localized
4$f$ electrons.
PrPb$_3$ has physical properties appropriate for this purpose: (1) a
simple cubic structure (space group $Pm\bar{3}m$), (2) a well-defined
CEF level scheme with small excitation energies, (3) a nonmagnetic
$\Gamma_3$ CEF ground state (GS) in a CEF potential with $O_h$ symmetry. 
In $^{141}$Pr-based compounds with a nonmagnetic GS, a strong
intra-ionic hyperfine coupling between the $^{141}$Pr nuclear spin
$\bi{I}$ and the 4$f$ electrons causes an effective nuclear dipolar moment
typically $10\sim 100$ times larger than the bare nuclear value. 
This enables one to monitor the 4$f$ electronic state in zero applied field
(ZF) by observing the local field caused by the hyperfine-enhanced (HE)
nuclear dipolar moment. 
Here we report the first observation of spontaneous $\mu^+$ spin
precession with {\it harmonic frequencies} in the paramagnetic state,
which indicates a coupling between the $\mu^+$ spin $\bi{S}$
and $\bi{I}$ with Ising-like anisotropy.
We discuss the interplay between $\mu^+$ and the localized $4f$ electrons on
the basis of the anomalous $\mu^+$ spin precession and a strongly
anisotropic $\mu^+$ Knight shift.

 Single crystal samples of PrPb$_3$ were grown by the
Bridgeman method. 
The $\mu^+$SR measurements were carried out at the
M15 and M20B beamlines, TRIUMF, Canada, the $\pi$M3 beamline, PSI,
Switzerland, and the $\pi$A port, muon science laboratory (MSL), KEK,
Japan. The single crystals with (001) cleavage plane were
mounted on a sample holder. ZF-$\mu^+$SR measurements were performed
with the spin-polarized $\mu^+$ along the [001] direction.
For the transverse field (TF) $\mu^+$SR measurements, a magnetic field of
2~T was applied along the [001] direction with an initial $\mu^+$ 
spin polarization $\bi{P}(t=0)$ perpendicular to the [001] direction.

\begin{figure}
\includegraphics[scale =0.28]{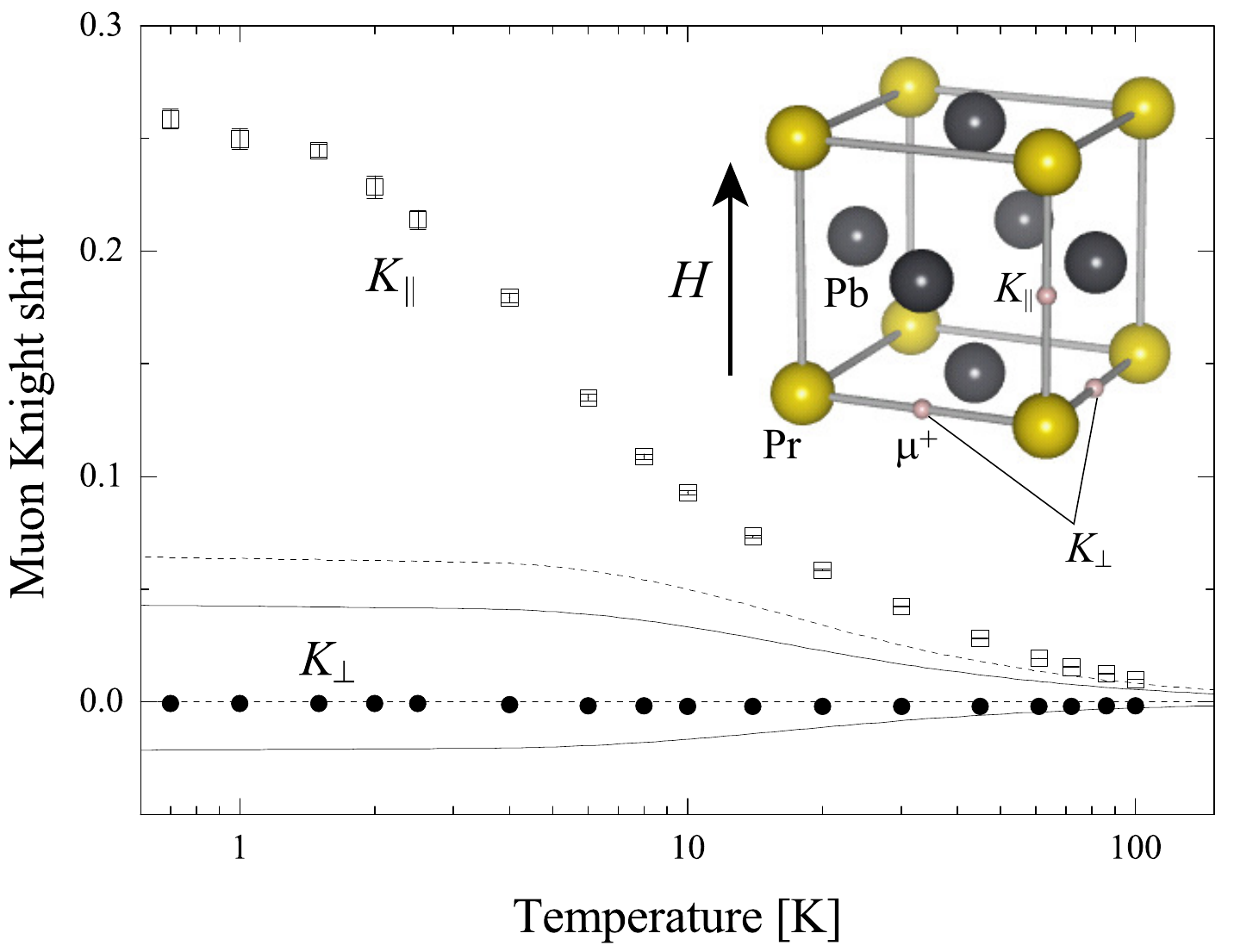}
  \caption{\label{Knight_Shift} The temperature dependence of the
  $\mu^+$ Knight shift in a magnetic field $\mu_0H=2$~T applied along the
  [001] direction. The inset shows the $\mu^+$ localization site.}
 \end{figure}

We first discuss the $\mu^+$ Knight shift, which provides insight
into the $\mu^+$ localization site, local susceptibility,
and the coupling between $\bi{S}$ and the 4$f$ dipolar moment.
The TF-$\mu^+$SR signal in the paramagnetic state consists of two
components with the intensity ratio of $1:2$, suggesting that the local
environment at the $\mu^+$ site splits into two magnetically
inequivalent locations. Among possible interstitial sites, only the 3$d$
site at (0.5, 0, 0) is consistent with this splitting. 
This site is located at the midpoint of two nearest neighbor (nn) Pr
ions, as depicted in the inset of Fig.~\ref{Knight_Shift}. The 1/3 (2/3)
signal with a frequency $f_{\parallel}$ ($f_{\perp}$) corresponds to the
site with four-fold symmetry around the [001] axis ([100] and [010] axes).

The $\mu^+$ Knight shift $K_{\parallel,\perp}$ was derived from
$K_{\parallel,\perp} = (f_{\parallel,\perp}-f_{0})/{f_{0}} -
K_{L,dem}$, where $f_0$ is a zero-shift frequency, $K_{L,dem}$ is a
correction term for the Lorentz and demagnetization fields (see
a preliminary report~\cite{tui07:_prpb3_knight_shift} for details).
Figure~\ref{Knight_Shift} shows the extremely anisotropic nature of the
shifts $K_{\parallel}\gg K_{\perp}$.
We now rewrite the Knight shift as a rank-2 tensor $\hat{K}$.
Hereafter, coordinate axes are set to be along primitive
vectors of the simple cubic lattice so that $\hat{K}$ becomes a
diagonal tensor. When the $\mu^+$ is located at the
site with four-fold symmetry around the $z$ axis, the diagonal
components $K_{ii}$ conform to $K_{xx}=K_{yy}=K_{\perp}$ and
$K_{zz}=K_{\parallel}$. Following the treatment in
Ref.~\cite{tashma97:_prin3}, we decompose $\hat{K}$ as
\begin{equation}
\hat{K}=(\hat{A}_{\rm nn}^{dip}+\hat{A}^{c})\hat{\chi}_{\rm nn}+\hat{A}_{\rm
 1-nn}^{dip}\hat{\chi},
\label{def_chi}
\end{equation}
where $\hat{A}^{dip}$s are classical-dipolar-coupling
tensors, $\hat{A}^c$ is an isotropic contact-hyperfine-coupling tensor,
$\hat{\chi}$s are ionic susceptibilities, and subscripts nn
and 1-nn designate the nn Pr ions and
the remaining further Pr ions, respectively. We assume that the $\mu^+$
influences only the nn Pr ions, 
so $\hat{\chi}_{\rm 1-nn}$ is replaced by an intrinsic susceptibility
tensor $\hat{\chi}$ in Eq.(\ref{def_chi}). This assumption seems
reasonable since the $\mu^+$ is located further
away from the Pr ions than their ligands, except for the nn Pr ions.
The contact term for the 1-nn Pr ions was omitted
for simplicity.

When $\hat{\chi}_{\rm nn}$ is identical to $\hat{\chi}$,
Eq.(\ref{def_chi}) reduces to
$\hat{K}=(\hat{A}^{dip}+\hat{A}^{c})\hat{\chi}$, where
$\hat{A}^{dip}=\hat{A}^{dip}_{\rm nn}+\hat{A}^{dip}_{\rm 1-nn}$.
In such a case, $\hat{A}^c$ is the only unknown variable in Eq.(\ref{def_chi})
since $\hat{\chi}$ is known from Ref.~\cite{tayama01} and
$\hat{A}^{dip}$ is purely geometric and calculable.
We obtained the solid lines in Fig.~\ref{Knight_Shift} for $\hat{A}^c=0$
and the dashed lines for $\hat{A}^c$ set to satisfy $K_{\perp}\sim 0$.
The significant deviation from the observed $\hat{K}$ suggests a
breakdown of our assumptions, implying (1) $\mu^+$-induced modification in
$\hat{\chi}_{\rm nn}(\neq \hat{\chi})$, and/or (2) temperature-dependent and
anisotropic $\hat{A_c}$ which has been invoked to explain $\mu^+$
Knight shift in several $f$ electron
systems~\cite{schenck04:_ceb6,lorenzi:_hob2c2}. 
By only measuring $\hat{K}$, the dominant factor cannot be pinpointed
in the present case. Fortunately, it turns out that the following
ZF-$\mu$SR results provide crucial information to resolve this point.
 \begin{figure}
\includegraphics[scale =1.75]{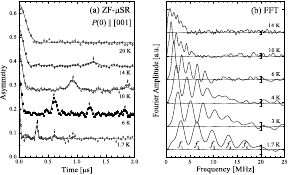}
  \caption{\label{ZF_Spectra} (a)ZF-$\mu^+$SR spectra above 1.7~K. The
  base lines are shifted by 0.1 as the temperature increases. The solid curves are the best fits with
  Eq.(\ref{eq_Pt_result}). (b)FFT spectra of the ZF-$\mu^+$SR
  signals (see text).}
 \end{figure}

\begin{figure}
 \includegraphics[scale =0.75]{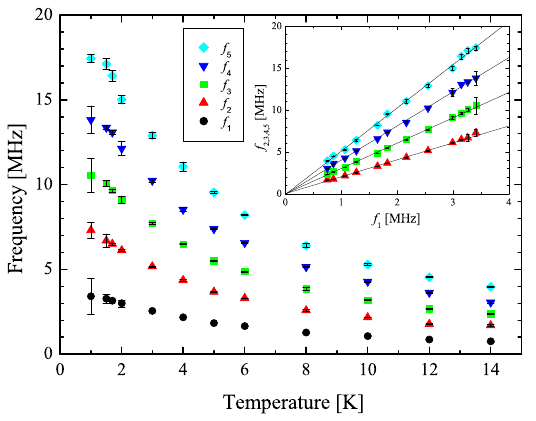}
  \caption{\label{ZF_Frequency} The spontaneous $\mu^+$ spin precession
  frequencies as functions of temperature. The inset is the $f_i$ versus
  $f_1$ plot ($i>1$). The solid lines represent the linear fits.}
 \end{figure}

The ZF-$\mu^+$SR spectra at several temperatures are shown
in Fig.~\ref{ZF_Spectra}(a).
Surprisingly, periodic recoveries of the asymmetry, indicated by the
arrows, are found below $T \lesssim 15$~K. 
Figure~\ref{ZF_Spectra}(b) shows results of fast Fourier transform (FFT)
on the ZF-$\mu^+$SR signals after subtraction of a zero-frequency
component. This reveals
that the original spectra consists of five oscillatory components (six,
when the zero-frequency component is counted).
The $\mu^+$ spin precession frequencies $f_i$ are shown in
Fig.~\ref{ZF_Frequency} as functions of temperature.
The frequencies monotonically increase with decreasing temperature, 
and the lowest frequency $f_1$
reaches $\sim$3.4~MHz at 1~K, which corresponds to a hyperfine field
(HF) of $\sim$0.025~T at the $\mu^+$ site, much larger than the bare nuclear
dipolar field. However, no magnetic ordering is expected since the
$\Gamma_3$ GS is nonmagnetic.
These facts indicate that the HF principally originates from the 4$f$
dipolar moments induced by the intra-ionic hyperfine interactions.
A linear relation between the $f_i$s and $f_1$ is shown in the inset of
Fig.~\ref{ZF_Frequency}. Slopes obtained from the linear
fits indicate that the $f_i$s are integral multiples of the $f_1$. This
means that the HF at the $\mu^+$ is {\it quantized}. 
The quantum nature and the number of
splittings imply a coupling between $\bi{S}$ with $S=1/2$ and a small number of
$\bi{I}$s with $I=5/2$.

The periodic structure in the ZF signals gradually disappears below
$\sim$ 1~K, probably as the temperature approaches $T_{Q}=0.4$~K, where an
antiferro-quadrupolar (AFQ) ordering with a long ordered structure
occurs~\cite{onimaru05}. 
This behavior may be related to the AFQ ordering and its precursive
phenomenon, which changes the HE nuclear magnetism through
modifications of the $4f$ electronic state.

We are now ready to construct a phenomenological model to describe the
ZF-$\mu^+$SR results.
It is known that a strong coupling between a $\mu^+$ spin and a small
number of nearby nuclear spins causes spontaneous $\mu^+$ spin
precession even in the paramagnetic state.
This has been observed in some fluorides in which an entangled
state of the nuclear spins, $^{19}$F-$\mu^+$-$^{19}$F, is
realized~\cite{brewer86:_fmf}. 
In the case of $\mu^+$ in PrPb$_3$, two Pr ions and an interstitial
$\mu^+$ form a collinear spin configuration
$^{141}$Pr-$\mu^+$-$^{141}$Pr. Therefore, we consider the following spin
Hamiltonian: $H = H_{SI}+H_{II}+H_{quad}$, where $H_{SI}$ and $H_{II}$ express spin interactions
between $\bi{S}$ and $\bi{I}$s, and between $\bi{I}$s, respectively, 
and $H_{quad}$ is a Hamiltonian for the nuclear quadrupolar interaction under
the electric field gradient created by the $\mu^+$. We hereafter
distinguish the two nn Pr with subscripts 1 and 2.

Considering a coupling with the HE nuclear dipolar
moment $\bi{\mu}_{hf}$, we first formulate $H_{SI}$ as
$-\gamma_{\mu}\hbar\bi{S}\cdot\hat{a}(\bi{\mu}_{hf,1}+\bi{\mu}_{hf,2})$,
where $\gamma_{\mu}$ is the muon gyromagnetic ratio (=2$\pi\times$135.53
MHz/T), and $\hat{a}$ is the hyperfine coupling tensor defined for each 
$\mu^+$-Pr coupling.
The $\bi{\mu}_{hf}$ is rewritten with the $^{141}$Pr Knight shift tensor
$\hat{K}_{I}$ as
$\gamma_{I}(1+\hat{K}_{I})\bi{I}\hbar$, where $\gamma_{I}$ is the
gyromagnetic ratio of the bare $^{141}$Pr nucleus.
Assuming that the principal value of the $\hat{K}_I$ is much larger
than 1, 
we obtain the following expression, 
\begin{equation}
H_{SI}\sim-\gamma_{\mu}\gamma_{I}\hbar^2 \bi{S}\cdot
 \hat{a}\hat{K}_{I}(\bi{I}_1+\bi{I}_2).
\label{eq_H_Imu_1}
\end{equation}
Although the above form includes an unknown tensor $\hat{a}\hat{K}_I$,
it will turn out to be approximately proportional to the $\mu^+$ Knight shift
$\hat{K}$. 
The $\hat{K}$ is dominated by the coupling to the two nn Pr ions, 
so $\hat{K}\sim 2\hat{a}\hat{\chi}_{\rm nn}$.
The $\hat{K}_{I}$ is proportional to $\hat{\chi}_{\rm nn}$ through the
following relation, 
\begin{equation}
\hat{K}_{I}= \frac{a_J}{g_J\mu_B\gamma_I\hbar}\hat{\chi}_{\rm nn},
\label{eq_KI_chi}
\end{equation}
where $a_J$ is the intra-ionic hyperfine coupling constant between
$\bi{I}$ and the total angular momentum $\bi{J}$ of the 4$f$ electrons,
and $g_J=4/5$ is the Lande $g$-factor.
Substituting $\hat{K}_I$ from Eq.(\ref{eq_KI_chi}) into 
Eq.(\ref{eq_H_Imu_1}) and using $\hat{K}\sim 2\hat{a}\hat{\chi}_{\rm nn}$
yields,
\begin{equation}
H_{SI}\sim-\frac{a_J\gamma_{\mu}\hbar}{2g_J\mu_B}\bi{S}\cdot\hat{K}(\bi{I}_1+\bi{I}_2).
\end{equation}
As shown in Fig.~\ref{Knight_Shift}, the $\hat{K}$ is strongly
anisotropic below the temperature at which the
spontaneous $\mu^+$ spin precession appears, and it is reasonable to set
$K_{\perp}=0$. 
Hence, Eq.(\ref{eq_H_Imu_1}) reduces to,
\begin{equation}
H_{SI}\sim-\hbar\omega_f S_{\parallel}(I_{1,\parallel}+I_{2,\parallel}),
\end{equation}
where $\omega_f=a_J\gamma_{\mu} K_{\parallel} / (2g_J\mu_{\rm B}$), and
$S_{\parallel}, I_{\parallel}$ are spin components parallel to the local
symmetry axis. This result implies an Ising-like anisotropy of the
$\bi{S}$-$\bi{I}$ coupling, which has not been reported so far in cubic
systems.

The $H_{II}$ and $H_{quad}$ should be invariant under the
symmetry operation of the system.
We will not discuss these terms further since they do not affect the
following outcome as long as this condition is satisfied.

With $\bi{P}(0)$ along the [001] direction, the time evolution
of the $\mu^+$ spin polarization $P(t)=\bi{P}(t)\cdot\bi{P}(0)$ is
expressed as follows, 
\begin{equation}
P(t) = \sum_{j=x,y,z} \frac{1}{3} {\rm Tr}\left[\rho e^{iH_j
					   t/\hbar}\sigma_z e^{-iH_j
					   t/\hbar}\right],
\label{eq_Pt}
\end{equation}
where $\sigma_z$ is the $z$ component of the Pauli spin matrices and 
$\rho=(1+\sigma_z)/[2(2I+1)^2]$ is the density matrix which describes
the spin state at $t=0$. 
It should be noted that explicit forms of the spin Hamiltonian $H_j$
depend on an angle between $\bi{P}(0)$ and the local symmetry axis.
The subscript $j$ is a label to indicate a spin
configuration with the local symmetry axis parallel to the $x$, $y$, $z$ axis. 

Calculating $P(t)$ according to Eq.(\ref{eq_Pt}) and multiplying it by
the total asymmetry $A$ yields the following function with oscillatory components
including {\it five harmonic frequencies} with a fundamental frequency
$\omega_f/2\pi$, i.e., 
\begin{align}
A&P(t) = \frac{A}{27} [ 5\cos(\omega_f t)+4\cos(2\omega_f t)+3\cos(3\omega_f
t) \nonumber \\
&+2\cos(4\omega_f t)+\cos(5\omega_f t) +3 ]e^{-\lambda_{\perp} t} +
 \frac{A}{3} e^{-\lambda_{\parallel} t}, 
\label{eq_Pt_result}
\end{align}
where the first (second) term corresponds to the spin configuration with the
symmetry axis perpendicular (parallel) to the $z$-axis.
Note that the exponential relaxation functions in
Eq.(\ref{eq_Pt_result}) with the relaxation rates
$\lambda_{\perp,\parallel}$ were inserted in the theoretical function to
express a transverse (longitudinal) relaxation for the first (second) term which
appears in the real spectra. Excellent fits are obtained for the
time-domain data below 14~K, as shown in Fig.~\ref{ZF_Spectra}(a) by the
solid curves. The relative amplitude of the oscillatory components shown
in Fig.~\ref{ZF_Spectra}(b) is also reproduced well.
On the other hand, if we assume $K_{\parallel}=2K_{\perp}$, which is
realized in the case of $\hat{\chi}_{\rm nn}=\hat{\chi}$ and
$\hat{A}_c=0$, more than five oscillatory components with
anharmonic frequencies should appear. This fact indicates that the 
Ising-like anisotropy is necessary to reproduce the harmonic frequencies
in the ZF-$\mu^+$SR spectra.

 \begin{figure}
 \includegraphics[scale =0.65]{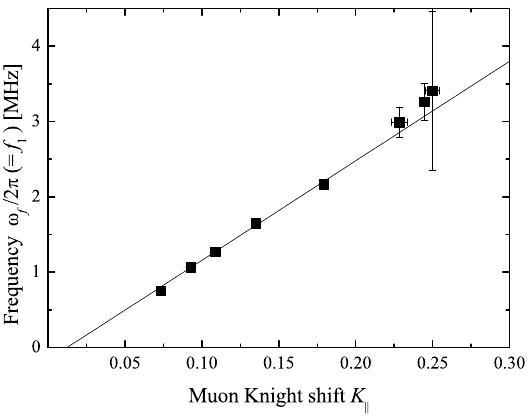}
  \caption{\label{F0vsK} The fundamental frequency $\omega_f/2\pi$
  ($=f_1$) versus the $K_{\parallel}$ in a field of
  2~T. The solid line indicates a linear fit.}
 \end{figure}

A value for $a_J$ is available from the present $\mu^+$SR data through
$a_J=2g_J\mu_{\rm B}/\gamma_{\mu}\times(\omega_f/K_{\parallel})$. The
ratio $\omega_f/K_{\parallel}$ was estimated to be 82.9 Mrad$/$s from the
linear slope in the $\omega_f/2\pi$ ($=f_1$) versus
$K_{\parallel}$ plot shown in 
Fig.~\ref{F0vsK}, so that we obtained $a_J/k_{\rm B}=105$~mK, which is
remarkably large compared to  the other trivalent RE ions~\cite{bleaney64}. The
present result thus provides microscopic evidence for the
HE nuclear magnetism of $^{141}$Pr.
It should be noted that our estimated value for $a_J$ is about two times
larger than that in Ref.~\cite{bleaney64}.
This discrepancy might be ascribed to an incomplete theoretical
formulation, where we used $\hat{K}_I\propto \hat{\chi}_{\rm nn}$, 
defined in a uniform field, to describe the HE nuclear dipolar moment in ZF.
In the uniform field, the Pr ion feels a molecular field from surrounding Pr
moments polarized along the field direction and $\hat{\chi}_{\rm
nn}$ should include the molecular field effect. On the other hand, the
Pr moments in ZF are randomly oriented and the effect of exchange interactions
among the Pr moments should be different from the molecular field in the 
uniform field.
The molecular field coefficient estimated from the magnetization is
-2.56~T/$\mu_{\rm B}$~\cite{tayama01} which causes suppression of
the field-induced dipolar moment, consistent with the overestimation of
$a_J$.

The observation of a spontaneous $\mu^+$ spin precession indicates
that the spin fluctuation rate $\tau_c^{-1}$ of the nn Pr ions is much
lower than 1~MHz.
By contrast, the intrinsic spin fluctuation rate
$|J_{nucl}|/h\sim$36~MHz is estimated from $T_c=|J_{nucl}|I(I+1)/3k_{\rm
B}$~\cite{ishida05:_prfe4p12}, with the nuclear ordering temperature
$T_c\sim$5~mK~\cite{abe03},
where $|J_{nucl}|$ is the nuclear exchange constant.
The marked reduction of $\tau_c^{-1}$
strongly suggests that the anisotropy in $\hat{K}$ is
principally due to that in $\hat{\chi}_{\rm nn}$, which suppresses the
$^{141}$Pr hyperfine-enhanced nuclear spin dynamics.
It should also be noted that previous arguments regarding the $\mu^+$-induced
perturbation on localized $f$ electronic states are based on the
significant deviation from $\hat{K}$-$\hat{\chi}$ scaling at low
temperatures~\cite{feyerherm:_prni5,tashma97:_prin3}. However, such a 
behavior also appears in NMR and $\mu^+$ Knight shifts in several
$f$ electron systems, and is ascribed to a temperature-dependent hyperfine
coupling~\cite{schenck04:_ceb6,lorenzi:_hob2c2,ohama95:_cecu2si2}, or to 
development of a local susceptibility attributed to the heavy electrons~\cite{ohishi07:_ce218,curro04:_kchi}.
 In this sense, an ambiguity exists in the interpretation of previous
 $\mu^+$SR data. The present observation thus 
provides the first experimental fact clearly illustrating that the localized
$f$ electronic state is modified by the local effects of an interstitial
charged particle. 

Finally, we discuss the $\mu^+$-induced change in the CEF level
scheme. Here we consider only low-lying $\Gamma_3$ and
$\Gamma_4$ states with an excitation energy
$\Delta_0=$14.7~K~\cite{tayama01} for
simplicity. The symmetry of CEF potential is lowered from cubic $O_h$ to
tetragonal $C_{4v}$ due to the $\mu^+$~\cite{tashma97:_prin3}. The
$\Gamma_3$ doublet and $\Gamma_4$ triplet split into nonmagnetic singlets
$\Gamma_3^{(1)}$ and $\Gamma_3^{(2)}$, and a $\Gamma_{4\pm}^{(1)}$
doublet and a $\Gamma_{4}^{(2)}$ singlet with magnetic degrees of
freedom, where we adopted the notation used in Ref.~\cite{tayama01}. The
low-temperature behavior in $\hat{K}$ indicates that the
magnetic anisotropy at low temperatures is dominated by the Van Vleck
term and a new CEF GS is still nonmagnetic. Since a non-zero matrix
element of $J_{\parallel}$ is found only between $\Gamma_3^{(1)}$ and
$\Gamma_{4}^{(2)}$ states, the modified level scheme should contain 
the $\Gamma_3^{(1)}$ singlet as the GS. Meanwhile, $\langle
\Gamma_3^{(1)}|J_{\perp}|\Gamma_{4\pm}^{(1)}\rangle \ne 0$. An 
excitation energy $\Delta$ of the $\Gamma_{4}^{(2)}$
($\Gamma_{4\pm}^{(1)}$) state must be smaller (larger) than $\Delta_0$
to reproduce the anisotropy in $\hat{\chi}_{\rm nn}$ since the Van Vleck
term is reciprocally dependent on $\Delta$. 
Furthermore, the large enhancement of $\hat{\chi}_{\rm nn,\parallel}$
implies that the $\mu^+$-induced energy shift is comparable in magnitude to
$\Delta_0$. This result suggests a possible drastic change of local magnetic
state around $^1$H impurities in RE-based intermetallics with
$\Delta\lesssim\Delta_0$.

We thank staff of the $\mu^+$SR facilities in TRIUMF, PSI, and KEK for
experimental assistance, as well as Profs. H.~Shiba, H.~Ishii, K.~Nishiyama, 
Drs. S.~Kambe, Y.~Tokunaga, H.~Sakai, H.~Chudo, and T.D.~Matsuda for
helpful discussions.
This work was supported by a Grant-in-Aid for Scientific Research on
 Priority Area ``Skutterudite'', Ministry of Education, Culture,
 Sports, Science and Technology, Japan, KEK-MSL 
Inter-University Program for Oversea Muon Facilities, 
and 21st Century COE Program at Tokyo Institute of Technology
``Nanometer-Scale Quantum Physics''.

\end{document}